%% file: main.tex
\documentclass[twocolumn,tighten]{aastex63}
\pdfoutput=1 
\usepackage{amsmath,amstext}
\usepackage[T1]{fontenc}
\usepackage{apjfonts} 
\usepackage[figure,figure*]{hypcap}
\usepackage{longtable}
\usepackage{threeparttable}

\usepackage{amssymb}
\usepackage{amsmath}
\usepackage{fontawesome}
\usepackage{gensymb}
\usepackage{mathrsfs}
\usepackage{upgreek}
\usepackage{fontenc}
\usepackage{hyperref}
\usepackage{float}
\usepackage{tabularx}
\usepackage{xcolor}

\renewcommand{\d}{\ensuremath{\rm d}}

\renewcommand{\textbf}{}

\shorttitle{An Accreting IMBH in a Metal Poor Dwarf?}

\graphicspath{{./}{figures/}}

\begin{document}

\title{Nuclear Activity in the Low Metallicity Dwarf Galaxy SDSS J0944$-$0038 : A Glimpse into the Primordial Universe}
\author[0000-0003-4701-8497]{Michael Reefe}
\altaffiliation{National Science Foundation, Graduate Research Fellow}
\affiliation{George Mason University, Department of Physics and Astronomy, MS3F3, 4400 University Drive, Fairfax, VA 22030, USA}
\affiliation{MIT Kavli Institute for Astrophysics and Space Research, Massachusetts Institute of Technology, Cambridge, MA 02139, USA}
\author[0000-0003-2277-2354]{Shobita Satyapal}
\affiliation{George Mason University, Department of Physics and Astronomy, MS3F3, 4400 University Drive, Fairfax, VA 22030, USA}

\author[0000-0003-3432-2094]{Remington O. Sexton}
\affiliation{George Mason University, Department of Physics and Astronomy, MS3F3, 4400 University Drive, Fairfax, VA 22030, USA}
\affiliation{U.S. Naval Observatory, 3450 Massachusetts Avenue NW, Washington, DC 20392-5420, USA}

\author[0000-0002-4902-8077]{Nathan J. Secrest}
\affiliation{U.S. Naval Observatory, 3450 Massachusetts Avenue NW, Washington, DC 20392-5420, USA}

\author[0000-0003-3937-562X]{William Matzko}
\affiliation{George Mason University, Department of Physics and Astronomy, MS3F3, 4400 University Drive, Fairfax, VA 22030, USA}

\author[0000-0002-6454-861X]{Emma Schwartzman}
\affiliation{George Mason University, Department of Physics and Astronomy, MS3F3, 4400 University Drive, Fairfax, VA 22030, USA}
\affiliation{U.S. Naval Research Laboratory, 4555 Overlook Ave. SW, Washington, DC 20375, USA}

\author[0000-0003-1991-370X]{Kristina Nyland}
\affiliation{U.S. Naval Research Laboratory, 4555 Overlook Ave. SW, Washington, DC 20375, USA}

\author[0000-0003-4693-6157]{Gabriela Canalizo}
\affiliation{Department of Physics and Astronomy, University of California, Riverside, 900 University Avenue, Riverside, CA 92521, USA}

\author[0000-0003-2283-2185]{Barry Rothberg}
\affiliation{George Mason University, Department of Physics and Astronomy, MS3F3, 4400 University Drive, Fairfax, VA 22030, USA}
\affiliation{LBT Observatory, University of Arizona, 933 N.Cherry Ave, Tucson AZ 85721, USA}

\author[0000-0001-8640-8522]{Ryan W. Pfeifle}
\altaffiliation{NASA Postdoctoral Program}
\affiliation{X-ray Astrophysics Laboratory, NASA Goddard Space Flight Center, Code 662, Greenbelt, MD 20771, USA}
\affiliation{Oak Ridge Associated Universities, NASA NPP Program, Oak Ridge, TN 37831, USA}

\author[0000-0003-1051-6564]{Jenna M. Cann}
\altaffiliation{NASA Postdoctoral Program}
\affiliation{X-ray Astrophysics Laboratory, NASA Goddard Space Flight Center, Code 662, Greenbelt, MD 20771, USA}
\affiliation{Oak Ridge Associated Universities, NASA NPP Program, Oak Ridge, TN 37831, USA}

\author[ 0000-0001-7578-2412]{Archana Aravindan}
\affiliation{Department of Physics and Astronomy, University of California, Riverside, 900 University Avenue, Riverside, CA 92521, USA}

\author{Camilo Vazquez}
\affiliation{George Mason University, Department of Physics and Astronomy, MS3F3, 4400 University Drive, Fairfax, VA 22030, USA}

\author[0000-0001-6812-7938]{Tracy Clarke}
\affiliation{U.S. Naval Research Laboratory, 4555 Overlook Ave. SW, Washington, DC 20375, USA}

\correspondingauthor{Shobita Satyapal}
\email{ssatyapa@gmu.edu}
\begin{abstract}

Local low metallicity dwarf galaxies are relics of the early universe and are thought to hold clues into the origins of supermassive black holes (SMBHs). While recent studies are uncovering a growing population of active galactic nuclei (AGNs) in dwarf galaxies, the vast majority reside in galaxies with solar or super solar metallicities and stellar masses comparable to the LMC. Using MUSE/VLT observations, we report the detection of [\ion{Fe}{10}] $\lambda$6374 coronal line emission and a broad H$\alpha$ line in the nucleus of  SDSS~J094401.87$-$003832.1, a nearby ($z=0.0049$) metal poor dwarf galaxy \textbf{almost 500} times less massive than the LMC. Unlike the emission from the lower ionization nebular lines, the [\ion{Fe}{10}] $\lambda$6374 emission is compact and centered on the brightest nuclear source,  with a spatial extent of $\approx$100~pc, similar to that seen in well-known AGNs. The [\ion{Fe}{10}] luminosity is  $\approx10^{37}$~erg~s$^{-1}$, within the range seen in previously identified AGNs in the dwarf galaxy population. The [\ion{Fe}{10}] emission has persisted over the roughly 19 year time period between the SDSS and MUSE observations, ruling out supernovae as the origin for the emission. The FWHM of the broad component of the H$\alpha$ line is $446 \pm 17$~km~s$^{-1}$ and its luminosity is $\approx~1.5\times10^{38}$~erg~s$^{-1}$, \textbf{corresponding to a black hole mass of $\approx~3150$~M$_\odot$, in line with its stellar mass if virial mass relations and black hole-galaxy scaling relations apply in this mass regime}. These observations, together with previously reported multi-wavelength observations, can most plausibly be explained by the presence of an accreting intermediate mass black hole in a primordial galaxy analog.


\end{abstract}

\keywords{galaxies: active --- galaxies: Starburst --- galaxies: Evolution --- galaxies: dwarf --- optical: ISM --- line: formation --- accretion, accretion disks }

\section{Introduction}

Intermediate mass black holes (IMBHs) in local dwarf galaxies, with masses predicted to be between $100 - 10^5$\,M$_\sun$ are crucial to our understanding of supermassive black hole (SMBH) seed formation and possibly dominate the extragalactic black hole population in the universe, yet black holes in this mass range have eluded detection by optical spectroscopic, mid-infrared color, X-ray, and radio surveys due either to obscuration of the central engine, or dilution of the accretion activity from star formation in the host galaxy \citep[see review by][]{greene2020}. Indeed, there is currently no black hole known to exist between approximately $150 - 10^4$\,M$_\sun$, a gap of roughly two orders of magnitude. Detecting black holes in this mass range kinematically is currently impossible, so a significant sample of IMBHs can therefore only be detected as active galactic nuclei (AGNs). Unfortunately, finding AGNs in dwarf galaxies is especially challenging, since the AGN luminosity is expected to be low, the star formation can be enhanced, and the gas-phase metallicities and black hole masses are expected to be low, all of which compromise the diagnostic potential of optical spectroscopic \citep{2006MNRAS.371.1559G, 2019ApJ...870L...2C}, X-ray \citep{2015ApJ...798...38S}, radio \citep{1991ApJ...378...65C, 2022ApJ...933..160S}, and broad-band mid-infrared \citep{2018ApJ...858...38S} surveys. Although there has been heroic work uncovering growing numbers of AGNs in the dwarf galaxy population \citep[see][and references therein]{greene2020}, most studies \textbf{are biased toward more massive black holes that reside in massive bulge-dominated and higher metallicity host galaxies, where contamination by star formation or obscuration by dust are less significant}. These traditional techniques for searching for AGNs are therefore severely limited in identifying the lowest mass black holes, and they have thus far not been able to identify black holes with masses less than $\approx10^4$\,M$_\sun$. Moreover, the vast majority of currently identified AGNs reside in galaxies with solar or super-solar metallicity, galaxies that do not resemble the primordial galaxies in the early universe. In these cases, the black holes cannot be considered relics of the seeds of supermassive black holes in the early universe \citep{2019NatAs...3....6M}. 

Observations of coronal lines are a promising path forward in the search for elusive AGNs missed by other widely used techniques \citep[see][and references therein]{2021ApJ...906...35S} and may even uncover and constrain the properties of accreting IMBHs \citep[e.g.,][]{2018ApJ...861..142C, 2021ApJ...912L...2C, 2022MNRAS.510.1010P}. These lines have already demonstrated the potential to uncover previously hidden candidate AGNs in bulgeless galaxies \citep{2007ApJ...663L...9S, 2009ApJ...704..439S} and the dwarf galaxy population \citep{2012MNRAS.427.1229I, 2021ApJ...911...70B,2021ApJ...912L...2C, 2021ApJ...922..155M, 2021MNRAS.508.2556I}, which are the likely hosts of IMBHs. Most importantly, coronal line activity can uncover accretion activity in the low metallicity galaxies where traditional optical narrow line diagnostics fail \citep{2019ApJ...870L...2C}.  In recent work, \citet{2022ApJ...936..140R} conducted the first systematic survey of a comprehensive set of the twenty optical coronal lines in the spectra of nearly 1 million galaxies observed by the Sloan Digital Sky Survey (SDSS), from the Data Release 8 catalog. The Coronal Line Activity Spectroscopic Survey (CLASS) remarkably revealed a population of coronal line emitters in dwarf galaxies ($M_* < 3 \times 10^9$\,M$_\sun$) that do not display optical narrow line ratios indicative of nuclear activity, many with metallicities well below those of any previously reported AGNs. This work also revealed that the emission lines with the highest ionization potential are preferentially found in lower mass galaxies, indicating the presence of hard ionizing radiation. Since lower mass black holes produce a hotter accretion disk \textbf{\citep{1973A&A....24..337S}}, this finding can be consistent with accreting IMBHs powering the coronal line emission in dwarf galaxies. Because of their low mass, low metallicity and compact morphologies, these galaxies are the best local analogs of the primordial galaxies thought to have formed at high redshift. In order to gain insight into the source of the hard radiation field and explore the diagnoistic potential of coronal lines in uncovering and constraining the properties of IMBHs, it is vital to carry out a multi-wavelength follow-up campaign of the population of coronal line emitters in the dwarf galaxy population.

In this work, we present follow-up archival integral field unit (IFU) observations with the Multi-Unit Spectroscopic Explorer (MUSE) on the Very Large Telescope (VLT) of SDSS~J094401.87$-$003832.1 (hereafter, J0944$-$0038), a nearby ($z=0.0049$) dwarf galaxy identified by CLASS with a [\ion{Fe}{10}] $\lambda$6374  detection in its SDSS spectrum. J0944$-$0038, also known as MCG+00-25-010, is a metal deficient blue compact galaxy that has previously been reported to display signatures of high ionization emission. It was reported to display a [\ion{Fe}{5}] $\lambda$4227 detection by \citet{2005ApJS..161..240T} and was also identified with an [\ion{Fe}{10}] $\lambda$6374  detection in the SDSS spectrum previously by \citet{2021ApJ...922..155M}.  Here we spatially map the emission lines and can constrain the location and spatial extent of the high ionization emission lines for the first time.  \textbf{The stellar metallicity of this dwarf galaxy is less than 10\%, and the stellar mass is approximately 500 times lower than the Large Magellanic Cloud (LMC), making it one of the lowest mass and lowest metallicity galaxies currently known in the universe to potentially host an AGN. Because of its low mass, low metallicity, and compact morphology, it belongs to a relatively rare class of chemically and possibly dynamically primitive local galaxies, with physical conditions characteristic of the early universe. Because of their proximity and the resulting potential for high sensitivity spectroscopy that is impossible to obtain at the very highest redshifts, these galaxies are ideal laboratories in which to gain insight into the first generations of stars, black holes, and the ionizing photons that contributed to cosmic reionization.} If current black hole galaxy scaling relations hold in this mass range, the central black hole will be significantly lower than any currently known. To estimate the distance to J0944$-$0038 we assume a flat $\Lambda$CDM cosmology with $H_0=70$~km~s$^{-1}$~Mpc$^{-1}$, and $\Omega_\mathrm{M} = 0.3$, and $\Omega_\Lambda = 0.7$. The resulting luminosity distance is $D_L=21$\,Mpc, consistent with that obtained using the Cosmicflows-3 local velocity flow model \citep{2019MNRAS.488.5438G,2020AJ....159...67K}.

\section{Target selection}

\begin{figure}
\includegraphics[width=\columnwidth]{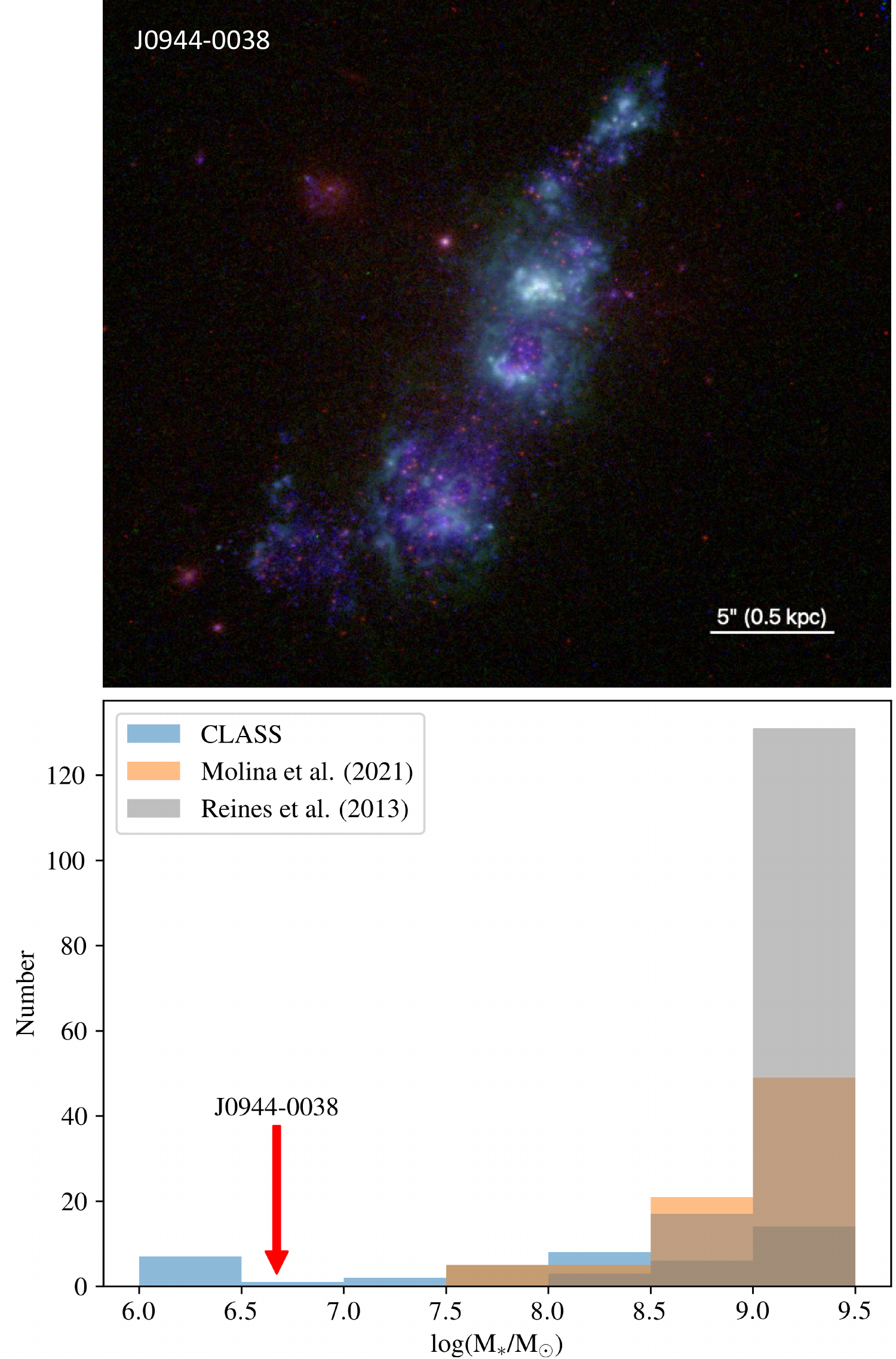}
\caption{Top panel: Hubble Space Telescope image of J0944$-$0038, constructed from UV (F336W, blue channel), H$\alpha$ (F657N, green channel), and visual/near-IR (F814W, red channel) archival data. 
Bottom panel: stellar mass distribution of the subset of dwarf galaxies in the CLASS sample compared to the samples of AGNs identified by \citet{2013ApJ...775..116R} and \citet{2021ApJ...922..155M}. The stellar mass of J0944$-$0038, and other coronal line emitters, is significantly lower than the mass regime probed by traditional optical narrow line ratios. }
\label{fig:mass}
\end{figure}

J0944$-$0038 is a dwarf galaxy with multiple star-forming knots dominated by two bright central optical continuum sources. It is a member of a poor cluster \citep{1984ApJ...283...33B} and is listed in Karachentsev’s catalog of isolated pairs of galaxies \citep{1972AISAO...7....3K}, with the brighter nucleus that coincides with the SDSS fiber listed as KPG~212A, and the fainter associated pair $1\farcm71$ south east is listed as 212B. A \textbf{$\approx~8\sigma$} [\ion{Fe}{10}] $\lambda$6374 detection is revealed in the SDSS spectrum but none of the other optical coronal lines were detected. \textbf{The total galaxy stellar mass obtained through broadband SED fitting of the UV + optical photometry using the new generation SED tool BEAGLE \citep{2016MNRAS.462.1415C} is $log M_* = {6.83}^{+0.44}_{-0.25}$ \citep{2022ApJS..261...31B}, almost 500 times less massive than the LMC.}
It is one of the lower mass dwarfs in the CLASS sample  and the sample of [\ion{Fe}{10}] emitters from \citet{2021ApJ...922..155M} with no optical evidence for an AGN based on its optical narrow emission line ratios, using the Baldwin-Phillips-Terlevich (BPT) diagnostics \citep{1981PASP...93....5B}, adopting the \citet{2003MNRAS.346.1055K} AGN selection criterion. As can be seen from Figure ~\ref{fig:mass}, the subset of dwarf galaxies that show coronal line emission from the CLASS survey and the survey by \citet{2021ApJ...922..155M} are uncovering a population of candidate AGNs in a significantly lower mass regime than identified in samples based on traditional optical narrow line ratios. The star formation rate based on the Balmer lines is $2.5\times10^{-2}$\,M$_\sun$\,yr$^{-1}$, and the stellar mass of \textbf{the brightest star forming clump}, based on fits to the broadband photometry, is $1.4\times10^5$\,M$_\sun$, assuming a constant star formation rate, or $2.2\times10^5$\,M$_\sun$  assuming a recent ``burst''\citep{2022ApJ...930..105S} .



 The gas-phase metallicity of the northern nucleus, based on SDSS and MMT observations is $12+\log(\mathrm{O}/\mathrm{H})$ = $7.81 \pm 0.05$ \citep{2017MNRAS.472.2608S}, placing it slightly above the 10\% solar metallicity cut employed to define the so-called ``Extremely Metal Poor (XMPs)'' galaxies. The metallicity of the stellar population of the northern nucleus, obtained through fitting the forest of UV photosheric absorption lines places is below 10\%, suggesting that the system is chemically similar to primordial high redshift galaxies \citep{2022ApJ...930..105S}. As a result, this system is an ideal probe of black hole seeds and the physics of massive stars and gas at very low metallicities.

\section{Observations and Data Reduction}

In this work, we employed MUSE \citep{2014Msngr.157...13B} archival data of J0944$-$0038, which was observed as part of a program to study starbursts in dwarf-dwarf interactions (Program ID: 0102.B-0325I; PI Privon). The observations were obtained on 2019~February~27 in the wide field  mode without adaptive optics (WFM-NOAO). The MUSE data covers the $475-935$\,nm wavelength interval with an average resolution $R\approx3027$ and a spatial sampling of $0\farcs2$ in WFM-NOAO. The effective exposure time for the observations was 2.65\,ks and the average seeing was $0\farcs686$. The processed MUSE data was retrieved from the ESO Archive Science Portal \footnote{https://archive.eso.org/scienceportal/home}, which had been processed using the MUSE 2.4 pipeline in which standard data reduction such as bias subtraction, flat-fielding, sky subtraction, wavelength and flux calibration, and removal of instrumental effects were performed. 

Spectral fitting was performed using using the open-source spectral analysis tool Bayesian AGN Decomposition Analysis for SDSS Spectra \citep[\textsc{badass};][]{sexton_2020} software, which has now been adapted to fit IFU data including MUSE.  The fitting procedure employs the  affine-invariant Markov-Chain Monte Carlo (MCMC) sampler \textit{emcee} \citep{2013PASP..125..306F} for robust parameter and uncertainty estimation.  The \textsc{badass} fitting routine uses the same procedure employed in the CLASS study in which the spectral continuum is modeled simultaneously with multiple components, including stellar continuum, AGN power-law, \ion{Fe}{2} emission, and spectral lines of various line profile shapes.  In order to enhance the signal-to-noise ratio (S/N) for detection of the coronal line emission, we employed the Voronoi binning algorithm VorBin \citep{2003MNRAS.342..345C}, which creates bins of constant S/N in the continuum. Fluxes from the MUSE data obtained using an SDSS-matched aperture were compared to the corresponding SDSS fluxes for all overlapping lines and the [\ion{Fe}{10}] fluxes were found to agree to within 16\%. 

In addition to fitting all strong lines and the [\ion{Fe}{10}] $\lambda$6374 line, which was previously detected by SDSS, we searched for all other coronal lines within the MUSE window to see if any new coronal lines were detected in the higher-sensitivity MUSE data. We also searched for broad lines in the MUSE cube by including broad components in the fits for detected lines. We utilize the line testing functionality built into \textsc{badass}, which performs comparison fits of the data with and without the line components in question and computes a detection confidence using \textbf{a custom likelihood-based model comparison test, the F-test, and reduced $\chi^2$ ratio}.
Using this method, we consider a strong detection to have a confidence $>95\%$.  We also test for the presence of lines using the pre-selection filtering algorithm BIFR\"OST, originally described in \citet{2022ApJ...936..140R}, \textbf{in which the average integrated flux in a region centered on the coronal line is compared to the root-mean-square (RMS) deviation of the flux in adjacent wavelength reference windows}.

\begin{figure*}[h]
\centering
\includegraphics[width=0.95\textwidth]{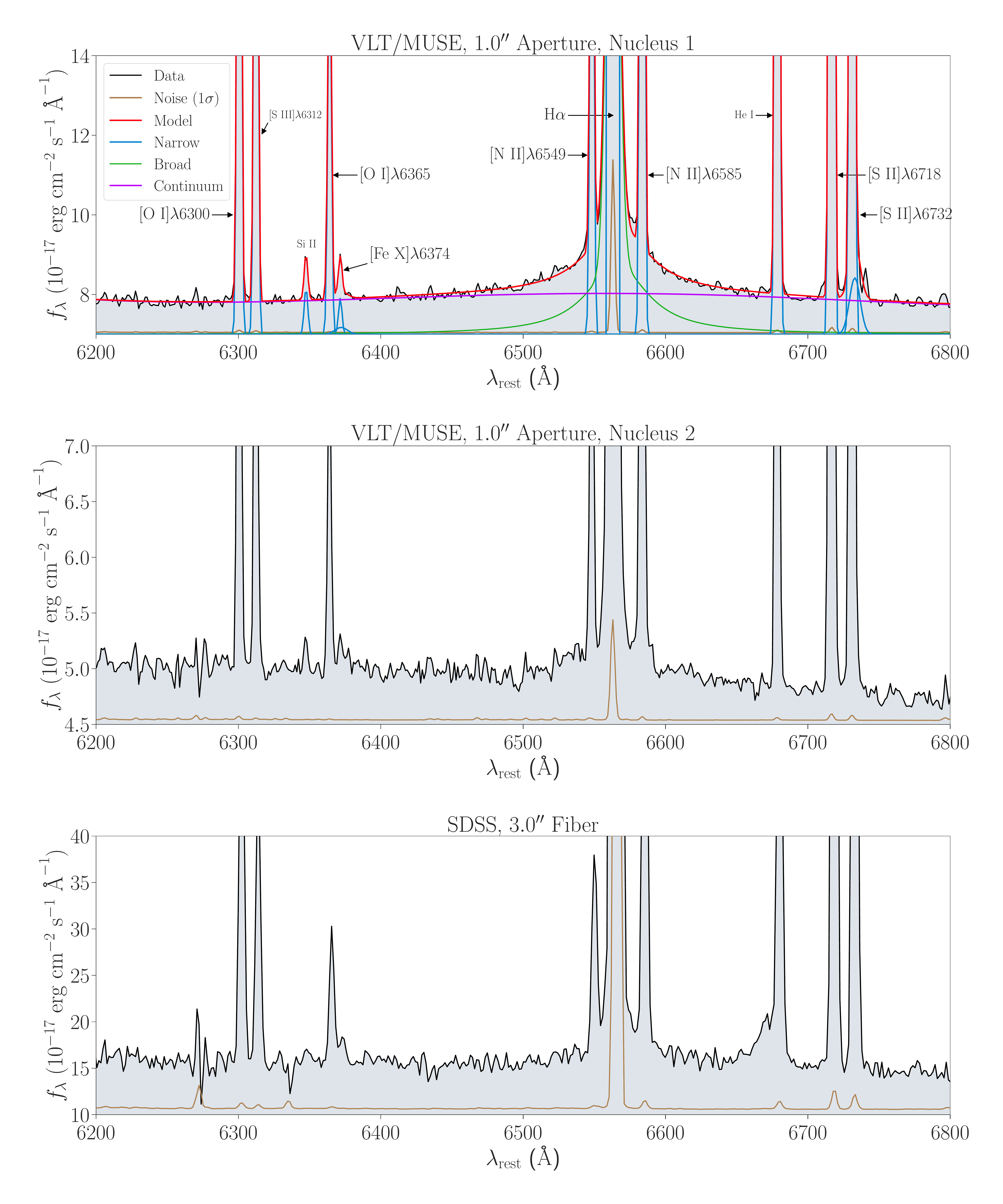}
\caption{
\textit{Top:} The 1D extraction of the MUSE spectrum from a $1\arcsec$ diameter circular aperture centered on Nucleus 1 (corresponding to the positions of the northern white circle in Figure~\ref{fig:musemaps}) showing the [\ion{Fe}{10}]\,$\lambda$6374 and H$\alpha$ region of the spectrum. The spectrum for Nucleus 1 was decomposed into broad and narrow components. A broad line was detected at the $>99.92~\%$ confidence level in Nucleus 1 with FWHM of $446 \pm 17$~km~s$^{-1}$.  \textit{Center}: The 1D MUSE spectrum from a $1\arcsec$ circular aperture centered on Nucleus 2 (southern white circle in Figure~\ref{fig:musemaps}).  \textit{Bottom}: The 1D SDSS spectrum from the $3\arcsec$ fiber.  No significant broad lines could be fit from the MUSE Nucleus 2 or SDSS spectrum.
}
\label{fig:spectra}
\end{figure*}

\textbf{In Figure~\ref{fig:spectra}, we show the 1D extracted MUSE spectra centered on the [\ion{Fe}{10}]\,$\lambda$6374 wavelength region from  $1\arcsec$ apertures centered on the two brightest starforming clumps, which we refer to in this work as Nucleus~1, and Nucleus~2, respectively (see Figure~\ref{fig:musemaps}). Nucleus~1, which corresponds to the brightest starforming region and is coincident with the SDSS fiber, shows a clear $\approx~10\sigma$ detection of the [\ion{Fe}{10}] line. For comparison, we also show the SDSS fiber spectrum from Nucleus~1 from the same wavelength region, which shows the $\approx~4\sigma$ detection of the [\ion{Fe}{10}]. There is also a tentative detection in the MUSE data at the location of Nucleus~2 at the $2.8\sigma$ level. The sensitivity of the current data is insufficient to determine if there are two distinct [\ion{Fe}{10}] sources possibly suggestive of a dual AGN in this dwarf merger. In addition, we detect a clear broad H$\alpha$ line in Nucleus~1, which is not seen in the less sensitive SDSS spectrum obtained with a larger $3\arcsec$ aperture.}

In order to accurately fit the broad H$\alpha$ component detected in Nucleus 1, we first determine the full extent of the narrow H$\alpha$ component by fitting a 4-moment Gauss-Hermite line profile with free parameters for amplitude, width, velocity offset, and higher-order moments $h3$ and $h4$.  We find that the narrow line profile is well-modeled by a Gaussian ($h3=0.009$, $h4=-0.008$), and has a width nearly identical to the [\ion{N}{2}]$\lambda6585$ line.  We then mask the brightest 10 pixels of the H$\alpha$ core to ensure that the fit to the broad wings of the H$\alpha$ component are not biased by the much-brighter narrow component.  The full broad line profile requires two components for an optimal fit, which includes a simple Gaussian component for the core of the broad line and a Lorentzian component for the extended wings of the profile.  The extended wings of the broad H$\alpha$ profile extend beyond the [\ion{N}{2}]$\lambda\lambda6549,6585$ doublet emission on either side of H$\alpha$, as shown in Figure~\ref{fig:spectra}.  The full composite line profile, composed of the individual two broad components, is summed and \textsc{BADASS} computes the line parameters (flux, width, velocity, etc.).  The total flux of the narrow component of H$\alpha$ is then determined from the residuals of the fit to the broad line from the masked pixels.




\section{Results and Discussion}

\subsection{Coronal Line Detection and Morphology of the Ionized Gas}

\begin{figure*}[h]
\centering

\includegraphics[width=0.49\textwidth]{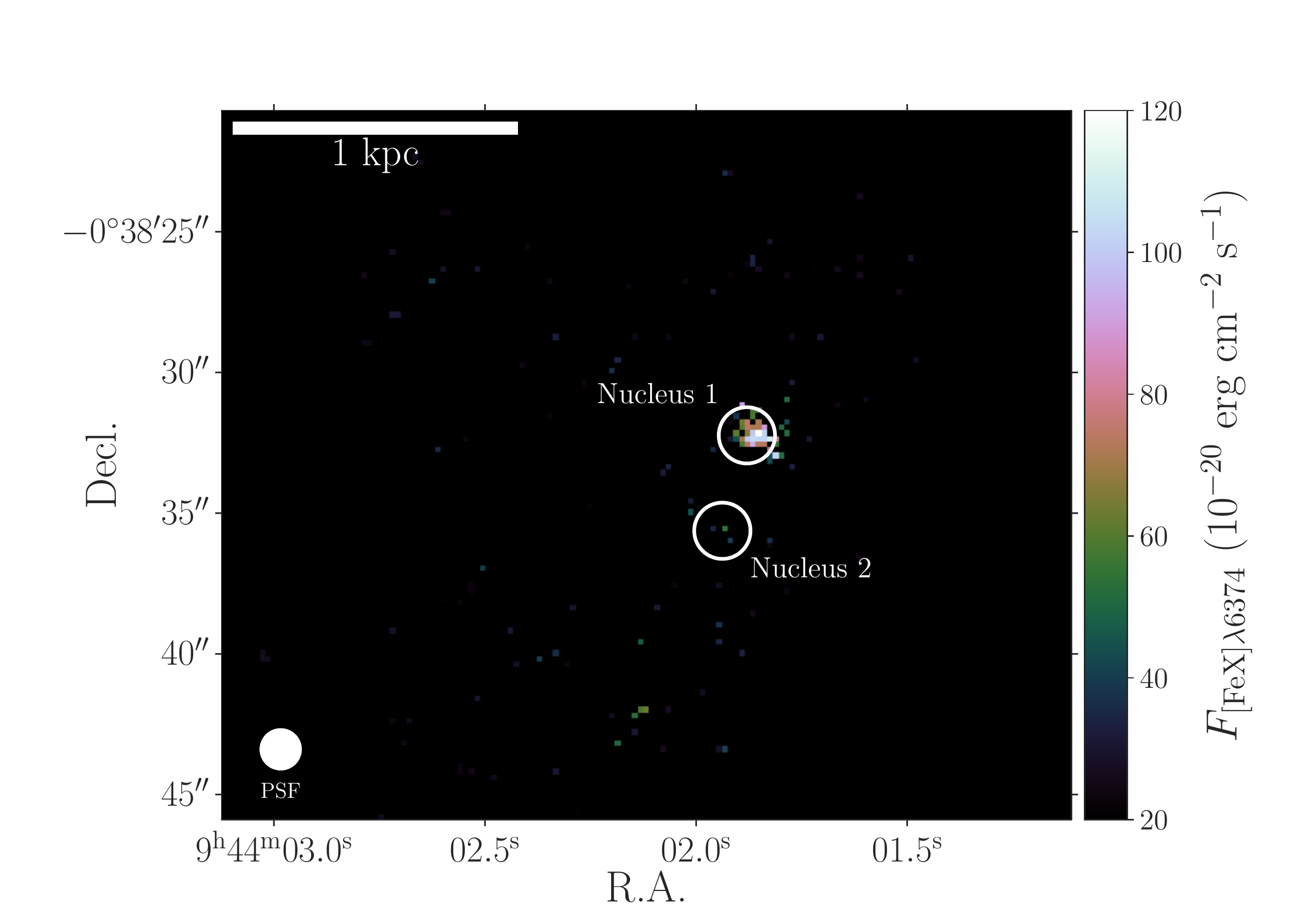} 
\includegraphics[width=0.49\textwidth]{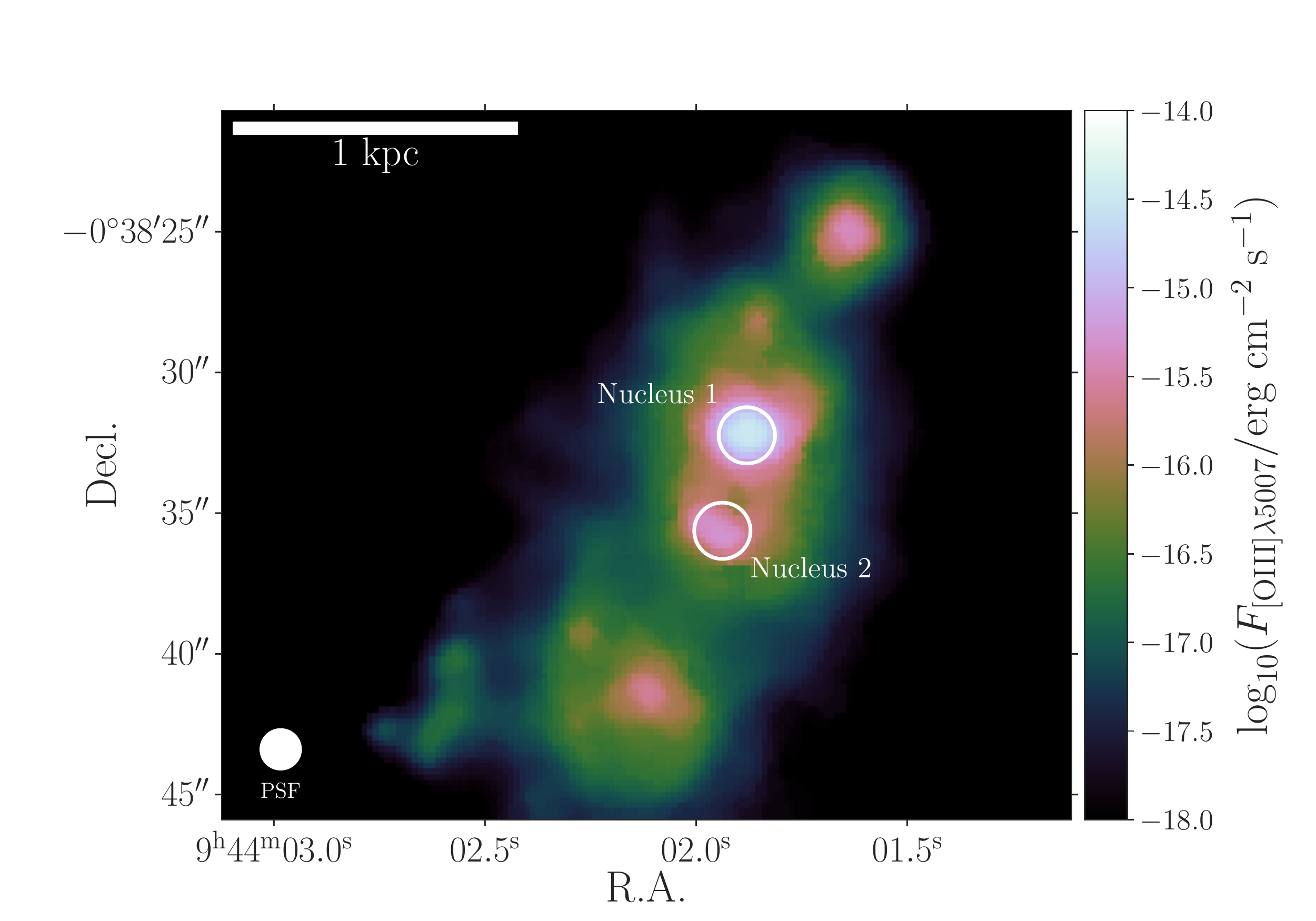}
\includegraphics[width=0.49\textwidth]{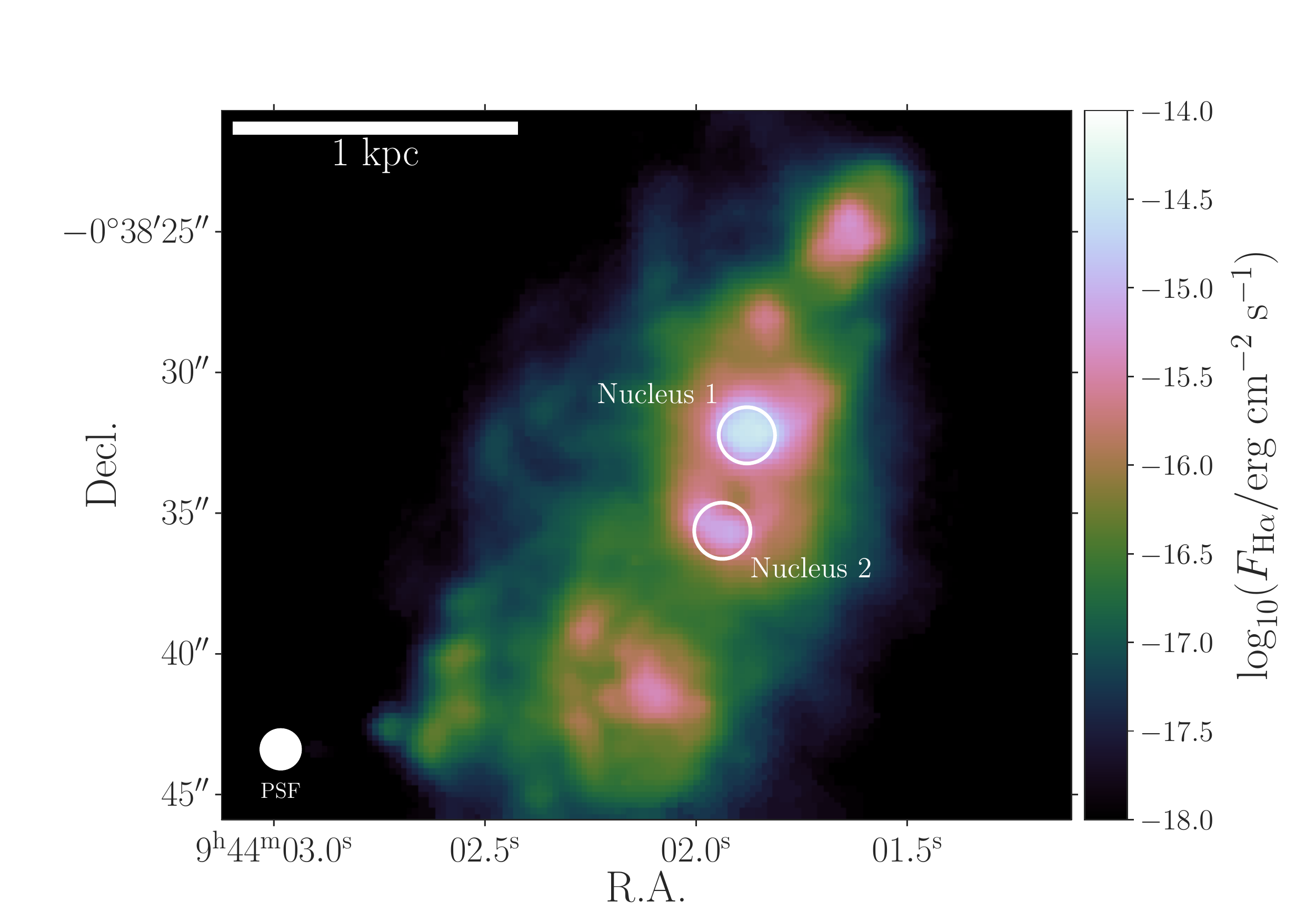} 
\includegraphics[width=0.49\textwidth]{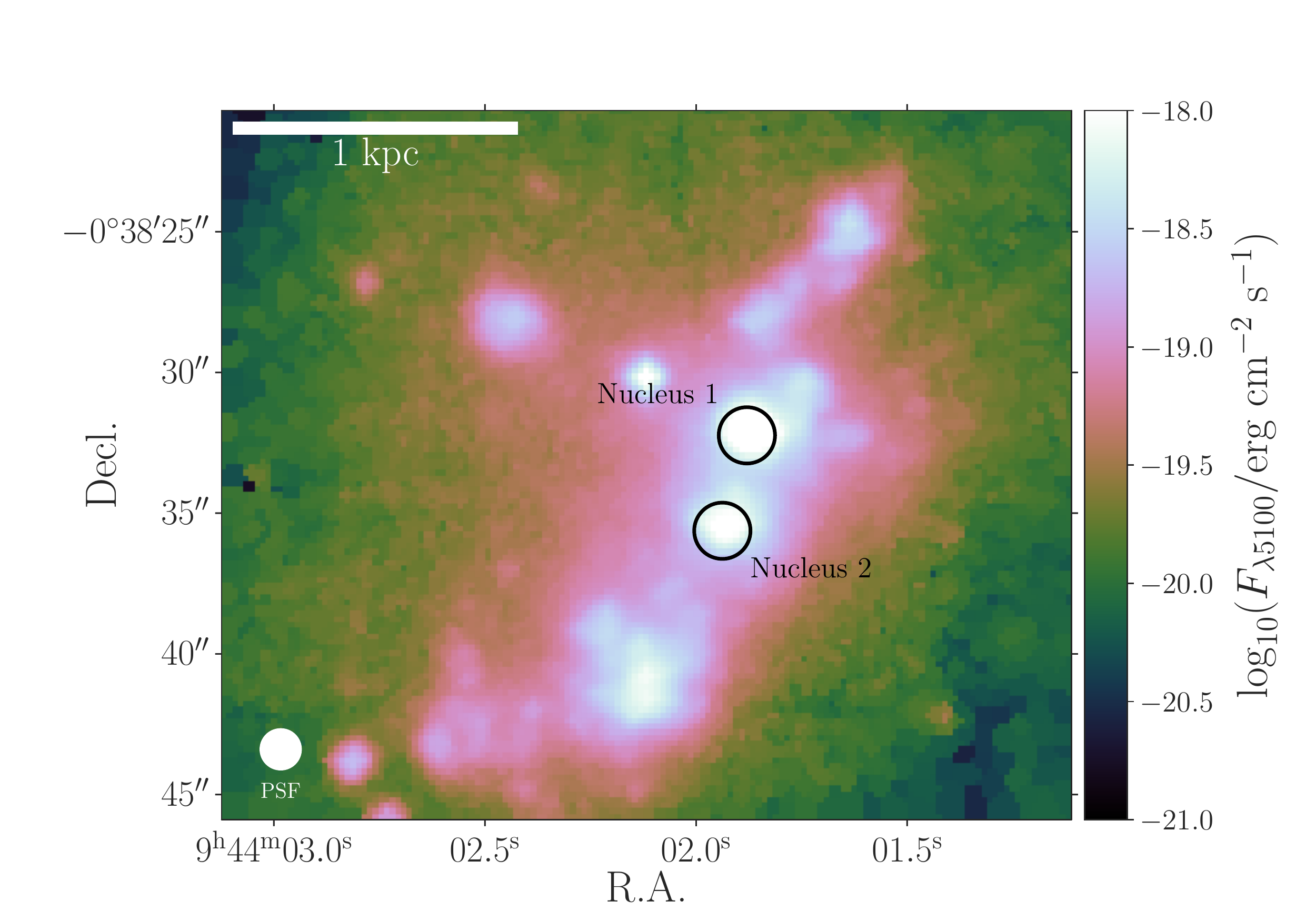}
\caption{\textit{Top Left:} The [\ion{Fe}{10}] $\lambda$6374 emission line flux map from VLT/MUSE. \textit{Top Right:} [\ion{O}{3}] $\lambda$5007 flux map. \textit{Bottom Left:} The H$\alpha$ flux map. \textit{Bottom Right:} The continuum at $\lambda$5100  flux map. The white circles in each image correspond to the positions of the peak continuum for the two brightest central clumps, which we refer to as Nucleus 1 (northern source), which is coincident with the SDSS fiber location, and Nucleus 2 (southern source). Scale bars in kpc is given in the top left of each plot.  The color scales for [\ion{O}{3}], H$\alpha$, and the continuum are logarithmic, while the color scale for [\ion{Fe}{10}] is linear. Additionally, the fluxes have been convolved with a 5x5 median filter, and [\ion{Fe}{10}], [\ion{O}{3}], and H$\alpha$ have been masked such that only spaxels with an S/N~$ > 3$ are shown.}
\label{fig:musemaps}
\end{figure*}

The [\ion{Fe}{10}] $\lambda$6374, together with the [\ion{O}{3}]\,$\lambda$5007, H$\alpha$, and continuum at $\lambda$5100  integrated flux maps are shown in Figure~\ref{fig:musemaps}. The galaxy shows four major star forming regions with diffuse emission and a shell-like structure seen connecting the two brightest continuum sources, \textbf{which we have referred to as Nucleus 1 and Nucleus 2 (denoted by white circles in in Figure~\ref{fig:musemaps}) }. While the [\ion{O}{3}]\,$\lambda$5007 and H$\alpha$ flux maps reveal similar morphologies, with emission from all star forming clumps and diffuse structures seen between the continuum sources, the  [\ion{Fe}{10}]\,$\lambda$6374 emission is compact and centered on the brightest source, Nucleus 1. The \textbf{[\ion{Fe}{10}] source is resolved} and the spatial extent of the emission detected above the $3\sigma$ level is $\approx 100$\,pc, comparable to the spatial extent of the highest ionization potential coronal line emission recently identified in nearby AGNs \citep[e.g.,][]{2011ApJ...739...69M, 2020ApJ...895L...9R, 2021MNRAS.506.3831F}. The total flux of the [\ion{Fe}{10}]\,$\lambda$6374 emission detected from the MUSE data is $2.82 \pm 0.48\times10^{-16}$\,erg\,cm$^{-2}$\,s$^{-1}$, which, using our adopted luminosity distance, corresponds to a luminosity of $\approx$ $10^{37}$\,erg\,s$^{-1}$, within the range of  [\ion{Fe}{10}]\,$\lambda$6374 luminosities observed in the dwarf AGN sample from \citet{2021ApJ...922..155M}. The ionization potential for [\ion{Fe}{10}] is 235\,eV, indicating extremely high energy photons from this source consistent with the presence of an AGN. Note that while coronal lines have been detected in planetary nebulae and Wolf-Rayet stars \citep[e.g.,][]{1999A&A...345L..17S}, their luminosities are several orders of magnitude lower, and no [\ion{Fe}{10}]\,$\lambda$6374 emission has been reported in Wolf–Rayet galaxies thus far \citep[e.g.,][]{2014MNRAS.439..157K}. We therefore rule out these stellar sources as the origin for the coronal line emission.

Supernovae (SN) have been reported to produce [\ion{Fe}{10}]\,$\lambda$6374 emission, but it is rare: the luminosities are typically orders of magnitude lower ($< 10^{33}$\,erg\,s$^{-1}$) than found in J0944$-$0038, and the emission fades considerably within the first few years \citep[e.g.,][]{1990AJ....100.1588B,2009ApJ...695.1334S, 2009ApJ...701..105K, 2017A&A...602L...4V, 2022MNRAS.515...71S}. The [\ion{Fe}{10}]\,$\lambda$6374 detection reported in this work is inconsistent with this scenario since the detection is seen both in the SDSS spectrum, which was obtained 2000~March~27, and the MUSE data, which was obtained almost 19 years later, with comparable flux in matched apertures given the photometric uncertainties. While the persistence of the coronal line emission over the roughly two decades between the SDSS and MUSE observations strongly disfavors SN activity as the origin of the emission, J0944$-$0038 was selected from an original search of $\sim1$~million galaxies for [\ion{Fe}{10}] emission, so we cannot rule out that an extremely rare luminous coronal line emitting SN such as 2005ip \citep[e.g.,][]{2009ApJ...695.1334S} happened to occur both in the 2000 SDSS spectrum epoch and in the 2019 MUSE spectrum epoch. To rule out this possibility, we extracted the MUSE spectrum from a $3\arcsec$ aperture centered on the [\ion{Fe}{10}] source, and compared it to the SDSS spectrum. Integrating over the SDSS~$r$ passband, we found that the SDSS and MUSE spectral fluxes differ from their mean value by 26\%, which is reasonably consistent with expected error in absolute flux calibration. The corresponding $r$-band AB magnitudes are 17.0 for SDSS and 17.5 for MUSE. These values correspond to absolute magnitudes of $\sim-14$ at the 21~Mpc distance of J0944$-$0038, far dimmer than the expected luminosities of any SN type \citep[e.g., Figure~1 in][]{2012Sci...337..927G}. While this finding alone argues strongly against a SN origin of the [\ion{Fe}{10}] emission, we compared these magnitudes to those from Pan-STARRS1 \citep[PS1;][]{2016arXiv161205560C} data obtained during a 4.1 year period between the SDSS and MUSE observations. At a mean epoch of 2012.8, the PSF $r$ mag of J0944$-$0038 from PS1 is 17.3, comparable to the MUSE and SDSS magnitudes, given typical calibration errors and errors induced by differences between aperture and PSF  magnitudes. Note that the optical continuum flux from supernovae in the $r$-band fades dramatically in the first few hundred days, typically by several magnitudes  \citep[e.g., see Figure~1 in][]{2009ApJ...695.1334S}. The consistency in magnitudes between the SDSS, PS1, and MUSE data therefore rule out SN activity as the origin of the [\ion{Fe}{10}] source. We note that a lack of detectable continuum variability in Nucleus 1 does not preclude the presence of an accreting IMBH. \citet{2020ApJ...900...56S} demonstrated that the relatively sparse temporal sampling of general survey data such as from the SDSS or WISE systematically misses AGN activity in dwarf galaxies. \textbf{In contrast to the dramatic and rapid variability seen in supernovae}, the fractional AGN variability on $\sim1$~yr timescales is typically about $\sim10\%$, which is below the sensitivity of the analysis performed here.

We searched for all optical coronal line detections but did not find additional detections within the MUSE passband, including the [\ion{Fe}{7}]\,$\lambda$6087 line, which is commonly detected in samples of larger mass galaxies \citep{10.1111/j.1365-2966.2009.14961.x}. This is however not unusual for the CLASS survey presented by \citet{2022ApJ...936..140R}. While both lines are often detected in higher mass galaxies, in lower mass galaxies, the higher ionization potential lines are more prominent. Of the the 72 galaxies with stellar mass $M_* > 3 \times 10^9$\,M$_\sun$ that show a [\ion{Fe}{10}]\,$\lambda$6374 detection, $\approx 68\%$ also show a [\ion{Fe}{7}]\,$\lambda$6087 detection. In contrast, only $\approx 17\%$ of galaxies with stellar mass $M_* < 3 \times 10^9$\,M$_\sun$ that show a [\ion{Fe}{10}]\,$\lambda$6374 detection, also show a [\ion{Fe}{7}]\,$\lambda$6087 detection. Of the 29 galaxies in the CLASS survey with $M_* < 1 \times 10^9$\,M$_\sun$ that display an [\ion{Fe}{10}]\,$\lambda$6374 line, not a single one shows the lower ionization [\ion{Fe}{7}]\,$\lambda$6087 line \citep[using catalog obtained from][]{reefe2022class}. These findings are consistent with a hardening of the radiation field with decreasing stellar mass, a result that would be expected with accretion onto lower mass black holes in the least massive dwarf galaxies. Such an effect is predicted by photoionization models with an accreting IMBH as shown by \citet{2018ApJ...861..142C}.

In Table \ref{tab:lines}, we list the fluxes obtained from the \textsc{badass} fits to the spectra from Nucleus~1 and 2, respectively. The flux ratio of [\ion{Fe}{10}] relative to the strong lines detected in the spectrum is somewhat lower than the mean values found in the CLASS sample (Reefe et al. 2023, in press), although there is a large dynamic range in the flux ratios across the CLASS sample, as well as in samples of well-known AGNs \citep[e.g.,][]{1997A&A...323..707E} or within the extended regions around single AGNs with IFU observations \citep[e.g.,][]{2021MNRAS.506.3831F}. The lack of detections of other coronal lines in this source is not unusual;  indeed, of the subset of BPT AGNs with coronal lines in the CLASS sample, $\sim$49\% show the detection of only one line.

\subsection{Broad Line Detection}

The FWHM of the broad H$\alpha$ line from the $1\arcsec$ diameter circular aperture centered on Nucleus 1 is $446 \pm 17$~km~s$^{-1}$. The total luminosity of the broad component is $\approx~1.5\times10^{38}$\,erg\,s$^{-1}$, comparable to that found in the well known type~1 dwarf AGN NGC~4395 \citep{2019MNRAS.486..691B}, and considerably lower than the broad line luminosities of previously identified low mass broad line AGNs in the dwarf galaxy population \citep[e.g.,][]{2007ApJ...670...92G, 2013ApJ...775..116R, 2022ApJ...937....7S}, as might be expected for a galaxy with a stellar mass at least one to three orders of magnitude lower than that of any previous dwarf galaxy with an identified broad line AGN \citep[e.g.,][]{2016ApJ...829...57B, 2022ApJ...937....7S}. Note that the broad line disappears when the extraction aperture size is increased to match the SDSS aperture size, indicating that its origin is from an extremely compact region centered on the nucleus. There was no evidence for a broad component or asymmetries in the profile of the [\ion{O}{3}]\,$\lambda$5007 line from this aperture, strongly suggesting that the broad H$\alpha$ line is consistent with virialized gas motion in a broad line region instead of outflowing gas. Note that while weak broad H$\alpha$ can be associated with supernova activity and has been \textbf{observed} to fade over time \citep[e.g.,][]{2016ApJ...829...57B}, the combination of the persistent [\ion{Fe}{10}] emission and lack of optical variability over almost two decades and the detection of a broad line in a compact aperture centered on Nucleus~1 is strong evidence for an AGN in this source. 

\begin{figure*}[h!]
\centering
\includegraphics[width=\columnwidth]{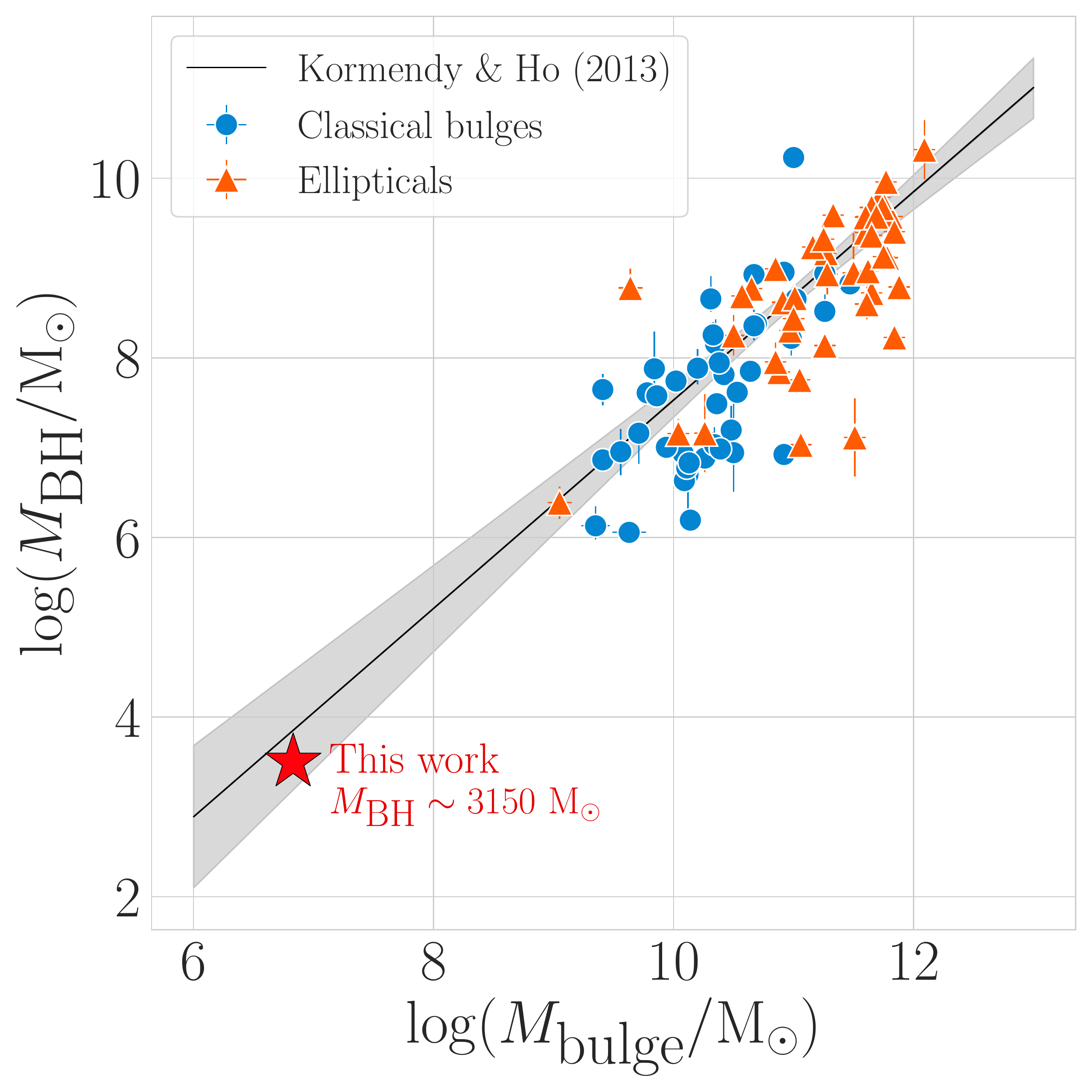}
\caption{The $M_\textrm{BH}$-$M_\textrm{bulge}$ relation for classical bulges and elliptical morphological types from \citet{2013ARA&A..51..511K} extrapolated to the estimated stellar mass of $6.76\times10^6$~M$_\odot$ and virial BH mass estimate of $M_\textrm{BH}\sim 3150~\textrm{M}_\odot$.  The shaded region represents the 95\% confidence interval for the fit to the $M_\textrm{BH}$-$M_\textrm{bulge}$ relation, which has an intrinsic scatter of 0.29 dex.}
\label{fig:kormendy_ho_2013}
\end{figure*}

\textbf{For the interests of curiosity, we approximated the black hole mass from the broad H$\alpha$ FWHM and luminosity using the virial mass relation from \citet{2012MNRAS.423..600S}, which is based on the calibrations from \citet{2005ApJ...630..122G} and \citet{2005ApJ...629...61K}.  Based on a FWHM of $446 \pm 17$~km~s$^{-1}$ and luminosity of $1.5\times10^{38}$~erg~s$^{-1}$, we estimate the black hole to have a mass of $\sim 3150$ M$_\odot$, however with some major caveats.  First, the calibration for H$\alpha$ virial black hole mass from reverberation mapped AGNs is performed for H$\alpha$ widths of $10^{3}$-$10^{4}$~km~s$^{-1}$ and luminosities between $10^{40}$-$10^{44}$~erg~s$^{-1}$, and thus our estimated mass is a linear extrapolation of the relation over nearly three orders of magnitude in black hole mass.  In addition to these extrapolations, the virial estimators are modeled assuming a strictly Gaussian width, whereas the best-fit model of our broad line is the sum of multiple gaussian and non-gaussian components, and given the strong dependence of width on virial mass, this could have a significant effect on our approximation.  Despite these important caveats, we plot our estimated black hole mass against the stellar mass for our object on the $M_\textrm{BH}$-$M_\textrm{bulge}$ for known AGNs, shown in Figure \ref{fig:kormendy_ho_2013} and find that it interestingly is consistent with the extrapolation of the scaling relations derived for more massive black holes. }

\medskip

In summary, the $\approx$ 10~$\sigma$ detection of the  [\ion{Fe}{10}] $\lambda$6374  line, which requires photons with energies $>$ 235~eV, its persistence over 19 years, the lack of optical variability in the SDSS, PS1, and MUSE data, its compact morphology centered on the peak continuum and the broad H$\alpha$ line identified in a compact aperture centered around Nucleus~1, all provide convincing evidence for the existence of an AGN in this low metallicity dwarf galaxy merger.

\input{line_table_letter}

\subsection{Other Evidence for an AGN}
There are several other lines of evidence that suggest the presence of a hard radiation field in J0944$-$0038 that can potentially be explained by the presence of an AGN . Based on the AllWISE catalog\footnote{\url{https://wise2.ipac.caltech.edu/docs/release/allwise/}}, the $W1-W2$ color, which includes both Nucleus 1 and 2, is $1.346  \pm 0.040 $, and has stayed consistently at this value over the nearly 12-year baseline of the WISE/NEOWISE multi-epoch catalogs. This $W1-W2$ color is  well above the widely used color cut ($>$0.8) used to identify AGNs from \citet{2012ApJ...753...30S}. In Figure~\ref{fig:bptwise}, we show the WISE colors of J0944$-$0038 along with several 3 band color cuts employed in the literature to identify AGNs \citep{2011ApJ...735..112J, 2018ApJ...858...38S, 2018MNRAS.478.3056B}. The mid-infrared colors do not meet the most stringent 3-band color cuts used to identify powerful AGNs, but its location in the mid-infrared color color diagram strongly suggests the presence of hot dust either heated by an extreme stellar population in a low metallicity galaxy or potentially heated by an AGN. 

While the optical narrow line ratios ratios from SDSS and the apertures centered on  Nucleus~1 and 2 are consistent with star formation (see Figure~\ref{fig:bptwise}), low metallicity galaxies with AGNs and AGNs powered by IMBHs are expected to show narrow line ratios that reside in the star forming region of standard BPT diagnostics \citep{2006MNRAS.371.1559G, 2019ApJ...870L...2C}.  Nucleus~1 displays prominent nebular emission in the resonant \ion{C}{4}\,$\lambda$1548,1550 doublet \citep{2022ApJ...930..105S}, exceedingly rare in nearby low redshift star forming galaxies but found in high redshift Ly$\alpha$ emitters \citep{2015MNRAS.454.1393S, 2017ApJ...836L..14M,2017ApJ...839...17S}. Nucleus~1 also displays a prominent \ion{He}{2}\,$\lambda$4686 line, as well as the $\lambda$1640 \ion{He}{2} lines in the UV \citep{2022ApJ...930..105S}, indicating the presence of highly ionized gas. While \ion{He}{2} emission, which requires the presence of ionizing photons with energies $>$ 54~eV, can be associated with Wolf-Rayet stars, the profiles produced by Wolf-Rayet stars are broad, in contrast to the line profile observed in this source. Based on the \textbf{lack of} stellar wind signatures  located near 4650\AA\ and 5808\AA, the galaxy is classified as a non-Wolf Rayet galaxy by SDSS, strongly suggesting that the observed \ion{He}{2} emission is not associated with Wolf-Rayet stars. Moreover, the \ion{He}{2}/H$\beta$ flux ratio is $0.0130 \pm 0.0006$ \citep{2017MNRAS.472.2608S}, well above that found in star forming galaxies and comparable to that found in \textbf{typical} AGNs \citep{2012MNRAS.421.1043S}. Because both lines are recombination lines with the same dependence on density and temperature, this line flux ratio is a strong indicator of the shape of the radiation field. The observed ratio in Nucleus~1 is too high to be explained by photoionization models with even the most extreme metal poor stars, and AGNs  have previously been invoked as an explanation in similar sources \citep{2018ApJ...859..164B, 2019MNRAS.490..978P}. Based on the models of \citet{2008ApJS..178...20A}, these authors point out that radiative shocks cannot reproduce the observed line ratios. In addition to the  [\ion{Fe}{10}] emission, J0944$-$0038 was one of 5 metal poor dwarf galaxies studied by \citet{2005ApJS..161..240T} to show [\ion{Fe}{5}] $\lambda$4227 emission. These authors argue that stellar population models cannot produce the required hard radiation field necessary to produce the emission. Based on the intensities of the near infrared  [\ion{Fe}{2}] lines, which are enhanced in shock excited gas, \citep{2016MNRAS.457...64I} argue that shocks are not necessary to explain the near-infrared spectrum of this source.

\begin{figure}
\includegraphics[width=\columnwidth]{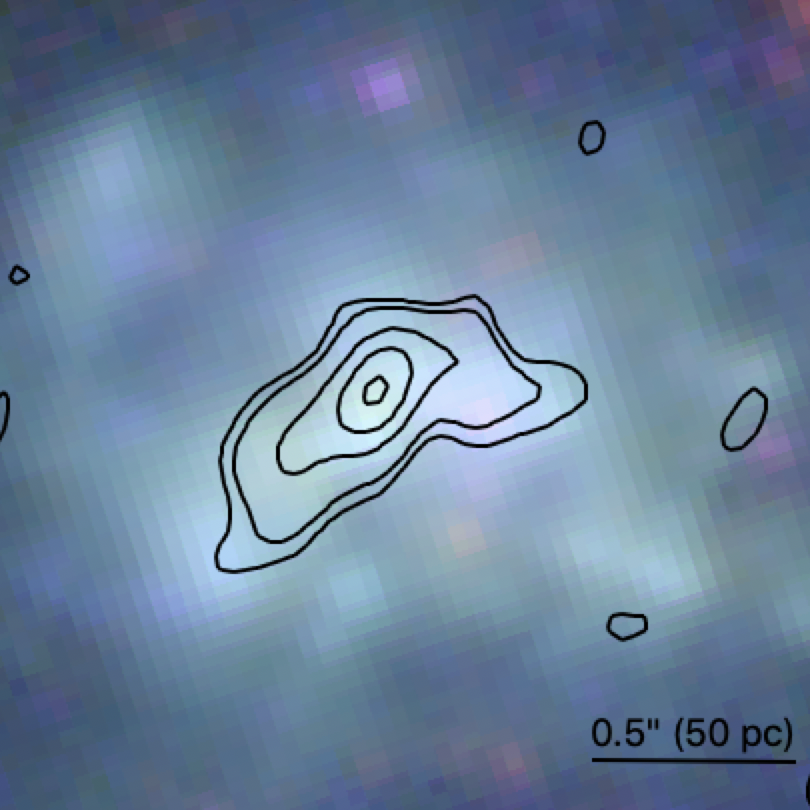}
\caption{VLA X-band contours of the radio emission coincident with the [\ion{Fe}{10}] source, overlaid on a $2\arcsec\times2\arcsec$ cutout of the HST image from Figure~\ref{fig:mass}.The contours start at the 3$\sigma$ level (23.5~$\mu$Jy~beam$^{-1}$) and increase by integer multiples of the $\sqrt{2}$. The VLA image has an angular resolution of 0.2$^{\prime \prime}$. }
\label{fig:vlaX}
\end{figure}

\begin{figure*}
    \centering
    \includegraphics[width=.9\textwidth]{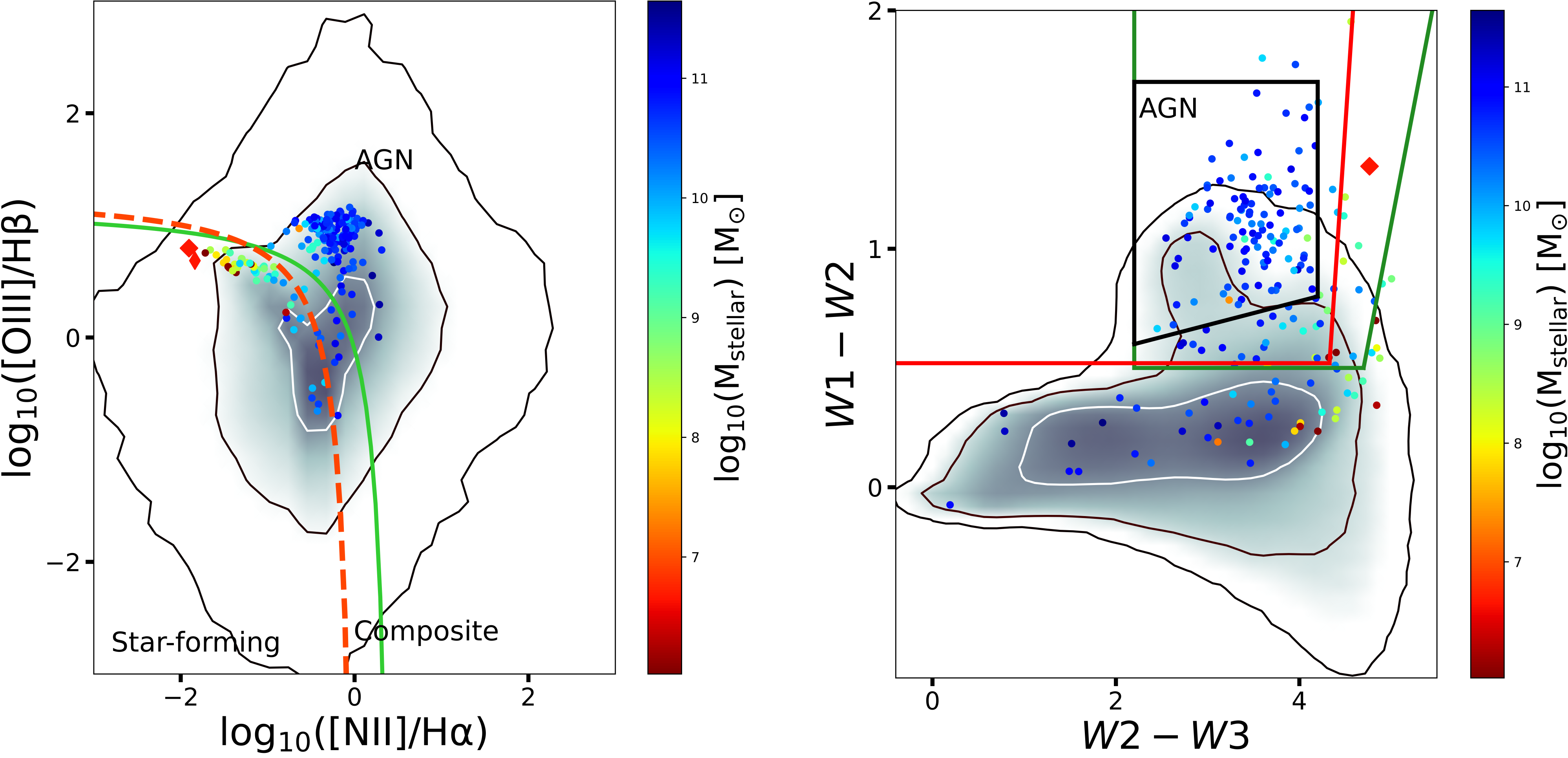}
\caption{\textit{Left}: A BPT ratio plot of the CLASS sample with the values obtained from the $1~\arcsec$ aperture centered on Nucleus 1 and 2 displayed by the red diamonds.  The color of each point corresponds to the stellar mass of the galaxy obtained from the MPA/JHU catalog.  The star-forming, ``composite,'' and AGN regions are marked with text and separated by the curves defined in \citet{2003MNRAS.346.1055K} and \citet{2001ApJ...556..121K}.  The grey shading displays the BPT values for the entire MPA/JHU catalog \\
    \textit{Right}: A WISE color-color plot of the class sample with the value for J0944$-$0038 displayed by the red diamond.  The coloring of the points and contours is defined the same way as the BPT plot.  The AGN demarcation box from \citet{2011ApJ...735..112J} is shown by the black line, the demarcation box from \citet{2018ApJ...858...38S} is shown by the red line, and the demarcation box from \citet{2018MNRAS.478.3056B} is shown in green. }
    \label{fig:bptwise}
\end{figure*}

Radio and X-ray emission can also be a valuable tool in detecting AGNs. J0944$-$0038 is one of the 0.03\% of dwarf galaxies in the sample by \citet{2013ApJ...775..116R} that show detectable radio emission. It is detected in the VLA Faint images of the Radio Sky at Twenty Centimeters (FIRST) survey \citep{1995ApJ...450..559B}. Follow up 10 GHz  VLA~A-configuration observations revealed an extended radio source centered on Nucleus~1 \citep{2020ApJ...888...36R}, as can be seen in Figure~\ref{fig:vlaX}. 
After estimating the contribution to the radio emission from \ion{H}{2} regions, supernovae, and supernova remnants using the global star formation rate estimated from GALEX and WISE, \citealt{2020ApJ...888...36R} concluded that the radio emission does not require the presence of an AGN. However, we note that there is considerable uncertainty in the relations between the star formation and the radio luminosity, and the radio morphology shown in Figure~\ref{fig:mass} may be consistent with star formation and a compact jet \citep{2017ApJ...845...50N}.

There is also a 15~ks Chandra observation of J0944$-$0038 (Obs~ID 19463; PI: Mezcua). These data were analyzed by \citet{2020MNRAS.494..941S} who report the detection of two prominent X-ray sources at the location of Nucleus~1 and 2, within positional uncertainties. 
\citet{2020MNRAS.494..941S} calculate the $0.5-8$\,keV flux assuming an absorbed power law model, with the column density set to the Galactic value toward the source, adopting a power law index of $\Gamma=1.7$. Using our adopted luminosity distance of 21~Mpc, the computed luminosity is $L_{\rm{0.5-8\,keV}}=4.9 \pm 2.1 \times10^{38}$\,erg\,s$^{-1}$, which is within the scatter of the predicted contribution from high mass X-ray binaries (HMXBs) given the star formation rate, using the recent relation for low metallicity galaxies from \citet{2022ApJ...930..135L}. Interestingly, Nucleus~2 is approximately an order of magnitude more luminous in the X-rays.  While there is no conclusive evidence for an AGN based on the radio and X-ray data alone, we point out that radio and X-ray observations are limited in the dwarf galaxy population. Indeed, only 0.3~\% of the sample of dwarf galaxies analyzed by \citet{2020ApJ...888...36R} were detected by FIRST. Of these, only 13 of 111 were selected as the most likely AGN candidates based on the follow-up VLA X-band A-configuration data, and a follow-up deep VLBA campaign \citep{2022ApJ...933..160S} demonstrated that four are likely background AGNs, with five remaining as possible AGNs despite non-detection with the VLBA. 
It is also challenging to use X-ray observations to identify AGNs in low mass galaxies since the X-ray contribution from HMXBs is significantly enhanced \citep{2022ApJ...930..105S} and recently confirmed AGNs in low metallicity dwarfs and low mass AGNs are found to be X-ray (and radio) weak \citep{2012ApJ...761...73D, 2016A&A...596A..64S, 2020ApJ...895..147C,2021MNRAS.504..543B}.  While inconclusive in confirming the existence of an AGN, the X-ray observations do place constraints on the HMXB and ULX population. \citet{2022ApJ...930..105S} demonstrate, using a grid of photoinization models, that photoionization by extremely metal poor stars coupled with multi-color disk models of HMXBs or ULXs cannot reproduce the observed \ion{He}{2}/H$\beta$ in this source. An additional source of ionizing photons consistent with an accreting more massive black hole is required to explain the nebular emission in this source.  

\subsection{Other Scenarios}

The presence and persistence of prominent  [\ion{Fe}{10}] emission in a low metallicity galaxy \textbf{almost 500 times} less massive than the LMC is surprising and difficult to explain without invoking the hard radiation field produced by an accreting IMBH or some other exotic physics. While other processes related to star formation activity have been invoked to explain nebular emission seen in the few dwarf galaxies that display  [\ion{Ne}{5}]\,$\lambda$3426 emission \citep[e.g.,][]{2005ApJS..161..240T, 2012MNRAS.427.1229I, 2021MNRAS.508.2556I}, the ionization potential of [\ion{Fe}{10}] is  235\,eV, considerably higher than the 97\,eV needed to produce [\ion{Ne}{5}]. If not explained by an AGN, the discovery reported here poses significant challenges to our understanding of the physics of extremely metal poor stars in low metallicity gas. 

The influence of massive stars on the nebular emission in J0944$-$0038 is clearly indicated by observations. P-Cygni profiles are seen in the \ion{C}{4} line profiles \citep{2022ApJ...930..105S}, indicating the presence of outflowing gas in the atmospheres of massive stars.  The possibility remains open that the physics of metal poor stars in low metallicity gas is not fully understood, and that \textbf{current models under-predict the emergent ionizing photon flux}. Unfortunately, stellar models in this metallicity regime have yet to be tested observationally, since samples of sub-solar individual stars only exist in the LMC and SMC, which have much higher metallicites than J0944$-$0038 (20\% and 50\% solar, respectively \citet{dufour_1984}).  Recent work on the only HII region observed that is powered by a single O star in an extremely poor (3\% solar) reveal no \ion{He}{2}\,$\lambda$4686 emission and a nebular emission line spectrum that is consistent with current metal poor stellar models, based on photoionization models \citep{2022arXiv221017535T}, suggesting that the observed emission line spectrum from J0944$-$0038 is inconsistent with a population of single metal poor hot stars. There have been recent studies suggesting that binary evolution, fast rotation, or other processes could lead to even hotter stars that could enhance the ionizing radiation field in low metallicity galaxies. For example, \citet{2019A&A...629A.134G} have shown that stars stripped of their envelopes from interaction with a binary companion, thought to be a common phenomena yet is neglected in stellar population models, produce a very hard radiation field that persists over longer evolutionary stages.   Inclusion  of binary stars and stellar rotation in the stellar synthesis models could significantly change the predicted ionizing radiation field and possibly account for the high ionization coronal line detected in low metallicity galaxies. Thus far, there has been no quantitative stellar population model that can account for nebular [\ion{Fe}{10} ], leaving open the possibility that the required hard radiation field is generated by as yet poorly understood physics in low metallicity stars in low metallicity galaxies, a result that is extremely important for our understanding of the first generation of stars in the early universe, and their role in reionization. 

\subsection{Metal Poor Dwarfs as Laboratories for the Early Universe}
\textbf{One of the longstanding goals of extragalactic astronomy is to understand the origins of the first stars, galaxies, and the supermassive black holes that reside in galaxy centers. The star formation rate, stellar mass growth, and accretion onto the first black holes in the early universe are also of critical importance in order to understand the first sources of ionizing photons and their contribution to cosmic reionization. While unprecedented sensitivity is now enabled by JWST in the high redshift universe, high sensitivity  of the weak coronal line spectrum will only be possible in the local universe. Because of their low metallicity, low mass, compact morphology, and high star formation rate, metal poor galaxies like J0944$-$0038 are the  best local analogs of primordial galaxies and ideal laboratories in which to study in detail the ionizing radiation field characteristic of galaxies in the early universe. Because of their high ionization potential, observations of the coronal lines can independently constrain the AGN contribution to the ionizing radiation field in contrast to the stronger lines from less ionized gas. Coronal lines therefore can  help provide constraints on the relative role of AGNs and star formation in reionization. In addition to being ideal laboratories in which to gain insight into the supermassive black hole seed population, metal poor dwarfs are ideal laboratories in which to gain insight into the physics of massive low metallicity stars, the primordial helium abundance, and the study of galactic chemical evolution over cosmic history. }
\textbf{}

\section{Summary and Implications}
In this work, using observations from MUSE/VLT, we report the detection of a [\ion{Fe}{10}]\,$\lambda$6374 line and a broad H$\alpha$ line seen only in a compact aperture centered on the brightest nuclear source in J0944$-$0038, a metal poor galaxy \textbf{almost 500 times} less massive than the LMC. This work highlights the potential of integral field observations of coronal line emission in identifying AGNs and constraining their location, and their power in finding faint broad lines missed in SDSS spectra. The persistence of the coronal line and lack of continuum variability over a 19 year period between the MUSE observations and observations from SDSS, together with the compact morphology, broad line width, and previously reported multiwavelength observations, strongly suggest that J0944$-$0038 harbors an accreting IMBH with a stellar metallicity and mass below typically reported in the dwarf galaxy population. If not explained by an AGN, these observations would imply the presence of an extreme stellar population, and that current stellar population models in the low metallicity regime significantly underpredict the ionizing photon flux, a result that will be extremely important in the interpretation of JWST spectra of high redshift galaxies. Since local metal poor galaxies like J0944$-$0038 can be considered analogs of primordial galaxies, a  true understanding of the ionizing radiation field in galaxies such as J0944$-$0038 is vital in our quest for understanding the sources of cosmic reionization.

\section{Acknowledgements}

M. R. gratefully acknowledges support from the National Science Foundation Graduate Research Fellowship under Grant No. 2141064. R.W.P. and J.M.C. would like to acknowledge a NASA Postdoctoral Program (NPP) fellowship at Goddard Space Flight Center, administered by ORAU through contract with NASA. G.C. and A.A. acknowledge support from the National Science Foundation under Grant No. AST 1817233.

\textbf{The authors would like to thank the anonymous referee for an extremely detailed and careful report that significantly improved the paper. The authors would also like to acknowledge helpful and insightful comments from Jonathan Stern, Ari Laor, Daniel Schaerer, and Marta Lorenzo. }

This research made use of Astropy,\footnote{\url{http://www.astropy.org}} a community-developed core Python package for Astronomy \citep{2013A&A...558A..33A}, as well as \textsc{topcat} \citep{2005ASPC..347...29T}.  


Funding for SDSS-III has been provided by the Alfred P. Sloan Foundation, the Participating Institutions, the National Science Foundation, and the U.S. Department of Energy Office of Science. The SDSS-III web site is \href{http://www.sdss3.org/}{http://www.sdss3.org/}.

SDSS-III is managed by the Astrophysical Research Consortium for the Participating Institutions of the SDSS-III Collaboration including the University of Arizona, the Brazilian Participation Group, Brookhaven National Laboratory, Carnegie Mellon University, University of Florida, the French Participation Group, the German Participation Group, Harvard University, the Instituto de Astrofisica de Canarias, the Michigan State/Notre Dame/JINA Participation Group, Johns Hopkins University, Lawrence Berkeley National Laboratory, Max Planck Institute for Astrophysics, Max Planck Institute for Extraterrestrial Physics, New Mexico State University, New York University, Ohio State University, Pennsylvania State University, University of Portsmouth, Princeton University, the Spanish Participation Group, University of Tokyo, University of Utah, Vanderbilt University, University of Virginia, University of Washington, and Yale University. 

This publication makes use of data products from the Wide-field Infrared Survey Explorer, which is a joint project of the University of California, Los Angeles, the Jet Propulsion Laboratory/California Institute of Technology, and NEOWISE, which is a project of the Jet Propulsion Laboratory/California Institute of Technology. WISE and NEOWISE are funded by the National Aeronautics and Space Administration.

The National Radio Astronomy Observatory is a facility of the National Science Foundation operated under cooperative agreement by Associated Universities, Inc.  Basic research in Radio Astronomy at the U.S. Naval Research Laboratory is supported by 6.1 Base Funding.  

\facilities{Sloan, VLT/MUSE, WISE}

\software{
Astropy \citep{2013A&A...558A..33A}, 
NumPy,\citep{2020Natur.585..357H}
SciPy,\citep{2020NatMe..17..261V}
\textit{emcee} \citep{2013PASP..125..306F},
pPXF \citep{2017MNRAS.466..798C},
\textsc{topcat} \citep{2005ASPC..347...29T},
\textsc{badass} \citep{sexton_2020},
BIFR\"{O}ST (\href{https://github.com/Michael-Reefe/bifrost}{https://github.com/Michael-Reefe/bifrost})
}

\nocite{1988AJ.....95...45A}
\nocite{2018ApJ...861..142C}


\bibliographystyle{yahapj}
\bibliography{main}







\end{document}

%% file: line_table_letter.tex
\begin{table*}[t]
\caption{Emission line fluxes}
\centering
\begin{tabular}{lccc}
\hline
\hline
\noalign{\smallskip}
    Line  & Wavelength & Nucleus 1 Flux\textsuperscript{1} & Nucleus 2 Flux\textsuperscript{1}\\
    & (\AA) & ($\log(F/\rm{erg}~\rm{cm}^{-2}~\rm{s}^{-1})$) & ($\log(F/\rm{erg}~\rm{cm}^{-2}~\rm{s}^{-1})$)\\

    \hline
    
    H$\beta$ & 4861 & $-13.547 \pm 0.008$ & $-14.121 \pm 0.004$\\

    [\ion{O}{3}] & 5007 & $-12.749 \pm 0.003$ & $-13.435 \pm 0.001$\\

    [\ion{Fe}{10}] & 6374 & $-16.41 \pm 0.04$ & $-17.07 \pm 0.15$ \\

    H$\alpha$ (narrow) & 6563 & $-12.887 \pm 0.003$ & $-13.486 \pm 0.001$\\

    H$\alpha$ (broad) & 6563 & $-14.534 \pm 0.025$ & ... \\
    
    [\ion{N}{2}] & 6585 & $-14.792 \pm 0.006$ & $-15.323 \pm 0.008$\\

    \noalign{\smallskip}
    \hline
    \noalign{\smallskip}       
     \end{tabular}
     \begin{tablenotes}

   \item[1] \textsuperscript{1}Fluxes obtained from BADASS fits from $1\arcsec$ circular aperture centered around Nucleus 1 and 2 (positions are displayed by white circles in Figure~\ref{fig:musemaps}).
\end{tablenotes}
     \label{tab:lines}
 \end{table*}

